# Detecting and Healing Area Coverages Holes in homogeneous Wireless Sensor Network: Survey

Baha Rababah, Ullash Saha

Department of Computer Science

University of Manitoba

Winnipeg, Canada

**Abstract:**

This paper presents Area coverage of homogenous wireless sensor network using computational geometry. The concepts related to both coverage wireless sensor network and computational geometry have been introduced. Then, the problem has been formulated. After that, current proposed algorithms of hole detection and healing have been investigated and discussed. This survey has two aims: firstly, to explore effectiveness of computational geometry to deal with wireless sensor network coverage problems. Secondly, it will direct scholars who want to enable computational geometry solutions in coverage Wireless sensor network problems.

**Index terms:** Wireless Sensor Network (WSN), Area Coverage, Computational Geometry, Coverage Hole Detection, Coverage Hole Healing.

## 1.0 Introduction:

Wireless sensor networks are a group of tiny nodes called sensors with embedded systems that able to senses, process and communicate with each other. In other words, WSN is a group of sensors work together as a team to achieve one or multi jobs such as monitoring and recording data at different locations. Wireless sensor networks become an essential part for wide range of technologies as it helps to monitor a wide range of parameters such as temperature, movements, rain, vibrations, gas concentrations, and others. However, WSN has many challenges such as bandwidth demand, cross layer design, and sensors deployment. Sensors should be able to deal with large volume of data in order to support current IoT applications, so deployment of sensors nodes should guarantee connectivity and coverage. WSN deployment can be defined as the method of locating the sensors in suitable positions for dealing with each other in productive way. Actually, the correct locations of sensors improve coverage and connectivity [1], it also extend the life time of the network. Coverage is to cover all the target area by sensors where the Euclidean distance between any point and the sensor covers its area is less than the sensing range [2]. Connectivity is that each pair of nodes is able to communicate directly or indirectly [3]. Coverage and Connectivity are two vital aspects to achieve acceptable level of quality of service in the WSNs. WSNs facing challenges in the topology nature, quantity of deployed sensors, deployment style, and sensing range. This paper is concerned in area coverage problem for homogeneous sensors and its proposed solutions with regard to computational geometry. The rest of the paper is represented as the following. Next section introduces related aspects to the topic. In the third section, presents area coverages and its categories. In the fourth section, problem formulation for area coverage of homogeneous wireless sensor is described. In the fifth and sixth sections, area coverage holes detections and healing algorithms are discussed. In the last two sections, challenges and conclusions are represented.

## 1.0    Related work:

Hole detection and recovery problem in WSN is widely studied in the field of computer science. Many approaches can be applied for finding and healing coverage hole. In this section we will only mention to the surveys that discussed area coverage detecting and healing, as we will mention to the current suggested algorithms in the hole detection and hole healing sections. In [4] Sangwan and Singh made a survey of k-coverage problem based on Voronoi diagram and Delaunay Triangulation. Yeasmin [5] also made a survey on k-coverage problem in WSN where at least k sensor nodes are required to cover an area. Similar type of surveys for coverage problem in WSN was done in [6, 7, 8] . This survey is the first survey that studies a certain problem in area coverage WSN which is area coverage for static homogeneous WSN Under computational geometry algorithms only. Computer geometry is one of the most popular approaches for the detection and recovery of coverage hole that cannot be ignored.

## 2.0 Related Aspects:

### 2.1 Computational Geometry:

Computational geometry is a field of computer science appear in 1970s to study of algorithms based to geometry [9].  it plays a basic role in improving many subjects such as computer graphics, computer and communication networks, Geographical Information system. In this survey we will mainly take look to two basic data structures belong to computational geometry: Voronoi diagram, Delaunay Triangulation.

### 2.1.1 Voronoi Diagram:

Voronoi diagram is splitting a plane of set of points into cells based on distances between the set of generating points. Each cell has one point from the set where all points in a certain cell are closer to this point than any other. Suppose we have a set of points P= { P1, P2,……, Pn} , P can be divided into several cells V(P) ={V(P1),

V(P2),…… V(Pn) } where every point in a certain region is closer to a certain point Pi than any other points in P. V(Pi)= {q/d(q,pi) < d(q,pj), i≠j, j=1, 2, 3….n}

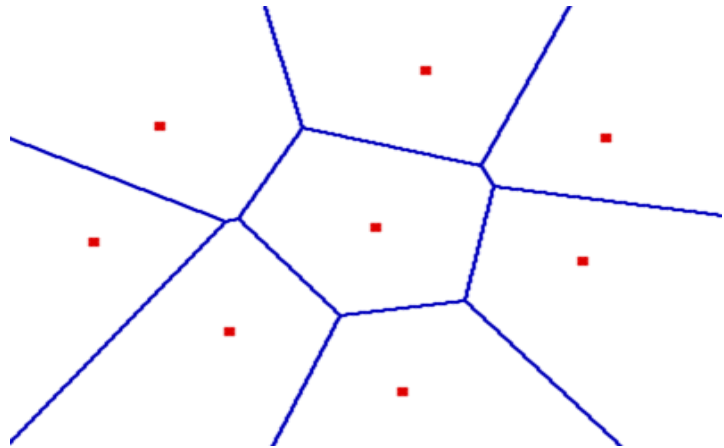

Figure 1 Voronoi Diagram

Voronoi diagram one of the essential concepts in computational geometry. Each adjacent cell separated by line called bisector which is perpendicular to the line segment. It uses in many algorithms for WSN coverage control. Figure 1 represents an example of using voronoi diagram for a set of points.

**2.1.2 Delaunay Triangulation:**

Delaunay triangulation for a given discrete set **P** of **n** points **{P1, P, ………,Pn}** in a plane is a type of computational geometry data structures that triangulate P where no points inside the circumcircle of any triangle . This algorithm plays an essential role in many coverage algorithms [10] [11] as it maximizes the minimum angle of all triangles in the plane [12].

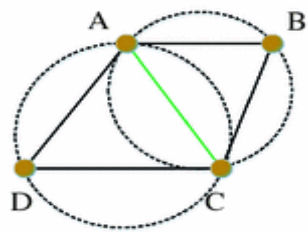

Figure 2 Delaunay Triangulation

**2.2 Related Definitions:**

**Sensing Range:** each sensor node can sense data within a range of radius $R_s$ .

**Communication Range:** Each sensor can communicate with other within its communication range of radius $R_c$ .

**Neighbour sensor node:** Two nodes (A and B) are neighbours to each other if the distance D between them is less than or equal twice the sensing range ($R_s$). $D_{A,B} \leq 2*R_s$.

**Boundary critical point:** A boundary critical point is the intersection point of a sensor node with its neighbour, which is not covered by any other sensor node.

**Hole boundary node:** The node which is closest to a hole is called border node or hole boundary node.

## 3.0 Area coverage:

The area of interest is covered by number of sensors, where each point of the field should be under the sensing range of one sensor. The networks cover the area may differ by types of sensors, deployment methods, or sensing and communication range.

3.1 Node Type: There are four types of Wireless sensor static, mobile, Homogeneous, and heterogeneous. The static node is fixed in a certain position, it is replaced when its energy drained. Many algorithms used static nodes are proposed [13]. Mobile node is able to change its position after initial deployment. The movement direction may be predefined or randomly based on the requirement and the aim of the algorithm [14]. Homogeneous means all nodes have similar capabilities of sensing, processing, storage and communication [15]. Heterogeneous means all nodes do not have the same capability of sensing, processing, storage and communication [15].

3.2 Deployment type: there are four main types of WSN deployment types classification based on different metrics such as network cost, network life time. Those types are deterministic, random, sparse, and dense. Deterministic means the node placement is already predetermined [16]. Random means the node position is selected randomly. Sparse is means to scatter nodes over an area. Dense means to deplot a large number of nodes in the area of interest.

3.3 Sensing and communication range: sensor network should be connected where any nodes able to reach the sink node. The relations between the sensing and communication range can be arbitrary or predefined such as Rc=Rs, *√3Rs ≥ Rc > Rs*, or *Rc ≥ √3Rs.*

## 4.0 Review of area coverage problem/ Problem formulation:

Let us assume that a set of n sensor nodes S= {s1, s2, s3, ....} are randomly scattered in a given region $R^d$ to perceive data. The region of interest $R^d$ can be convex or non-convex. All the sensor nodes are static. They are homogenous (same internal memory, power consumption, processing speed and so on). A sensor can perceive data of the area which is within its sensing range. The radius of sensing range is equal for all nodes ($R_{s1}=R_{s2}=R_{s3}=$ .........$=R_s$). A node can communicate with other nodes within its commination range. The radius of the communication range $R_c$ is equal for all nodes.

Problem 1: We need to detect the coverage holes (The areas which are not covered by any of the senor nodes) in the given area $R^d$ .

Problem 2: After the detection of the coverage holes, we need to cover the uncovered areas with minimum number of sensor nodes.

## 5.0 Hole Detection:

In [17, 18, 19] , some computational geometry approaches were used to detect coverage holes in WSN. In this paper we will highlight the recent works regarding computational geometry approach for detecting the holes in WSN.

Kang [20] explained that a coverage hole is formed in ROI when it is enclosed by continuous boundary critical points. The algorithm randomly selects a sensor node and finds its neighbour nodes. Then it finds out the critical points of the sensor node with its neighbours and with the boundary region. The algorithm detects the critical points for every sensor nodes through this procedure. Finally, it connects the continuous critical points and form a boundary line. The region enclosed by the boundary line is the hole area.

This algorithm is decentralized, co-ordinate free and node based. It uses the computational power of every node [21].

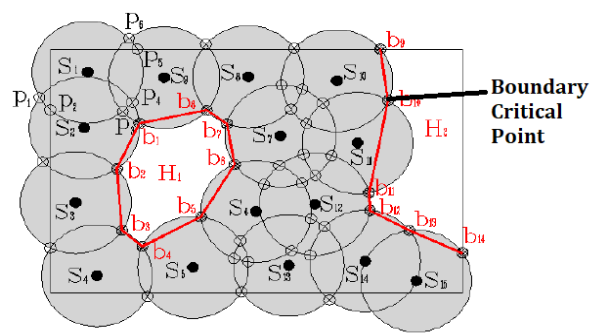

Figure 3 : Hole detection using boundary critical point. Adapted from [20]

Dai at el. [22] proposed an algorithm based on Voronoi diagram to detect the coverage holes. The algorithm finds the vertices of the corresponding Voronoi cell of each node. Then it calculates the distances between the node and each vertex of the corresponding Voronoi cell. The algorithm also computes the distance between the sensor node and each edge of its corresponding Voronoi cell. If any distance is greater than the radius of the sensor node then the node is tagged as hole boundary node.

The advantage of this algorithm is that it is easy to execute [22]. On the other hand, it only detects the boundary region of a hole but cannot measure the hole area [23]. Furthermore, all the nodes can communicate to each other to detect the hole and it requires massive power [23].

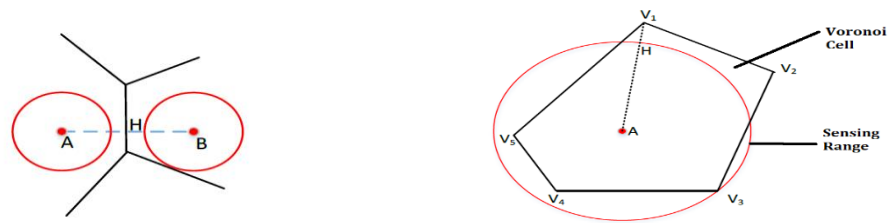

Figure 4 : Two cases of coverage hole in Voronoi cell. Adapted from [22]

Li and Wu [24] detected the coverage hole using the empty circle property of Delaunay triangle. According to the property, no node lies inside the empty circle. If the radius of the empty circle, $R_e$ is greater than the coverage range, $R_s$ of the sensor then there exists an uncovered area in that empty circle. The algorithm first builds the Delaunay triangulation from the given sensor nodes. Then it finds out the empty circles which radius is greater than the sensing range. In the next phase, the algorithm merges the selected isolated empty circles. There is a common side of two isolated empty circles that are generated from two neighbouring Delaunay triangles. If the length of the common side of two empty circles is greater than $2R_s$, they can be merged into one coverage hole. Even If the length of the common side is less than $2R_s$ but both centres of two empty circles lies on the same part of the common side, they are assigned to the same coverage hole. All the area of an empty circle ($R_e > R_s$) is not uncovered. Li and Wu initiated the concept of IEC which is a disc concentric with its corresponding empty circle and its radius is $R_{iec} = R_e - R_s$ . Each IEC of its corresponding empty circle is the estimated uncovered area of that empty circle.

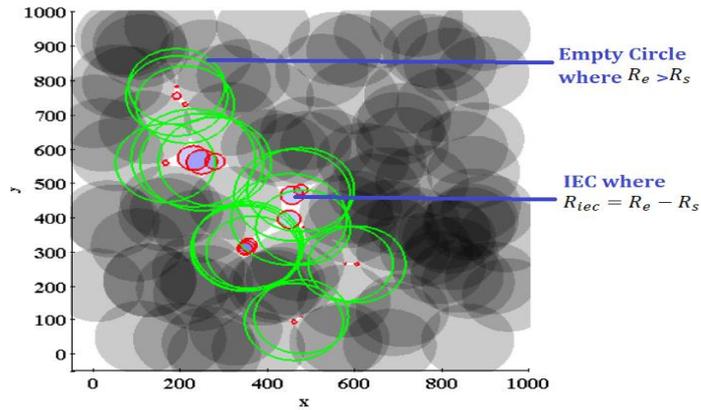

Figure 5: Area coverage hole detection using Delaunay Triangulation. Adapted from [24]

Koreim and Bayomi [23] proposed another algorithm(WHD) to detect the holes and calculate their areas in WSN. In the first phase, ROI is divided into many predetermined small regions of same size and shape by the grid algorithm [25]. Each small region is called a cell. In the second phase it identifies the closest sensor nodes of each cell. A node is chosen as the head node for every cell to communicate with other nodes. Then the algorithm compares sensor range of each closest node with each corner point of the corresponding cell. If any one of the corner point is not covered by any closest node then there exists a hole in that cell. After the detection of a hole in a cell, the algorithm computes the area of that hole using triangulation method between the uncovered edge point and all selected nearest nodes. The algorithm then simply calculates the area of each triangle and sums them to find the area of a single hole of a cell. Through this process the algorithm figures out the area of holes of every cell.

The main outcome of this proposed algorithm is reduction of energy consumption. Only one single head node is responsible for each cell to communicate with base station. Moreover, the collection of data of each head node is minimal as the number of nearest nodes for each cell is minimal [23].

.

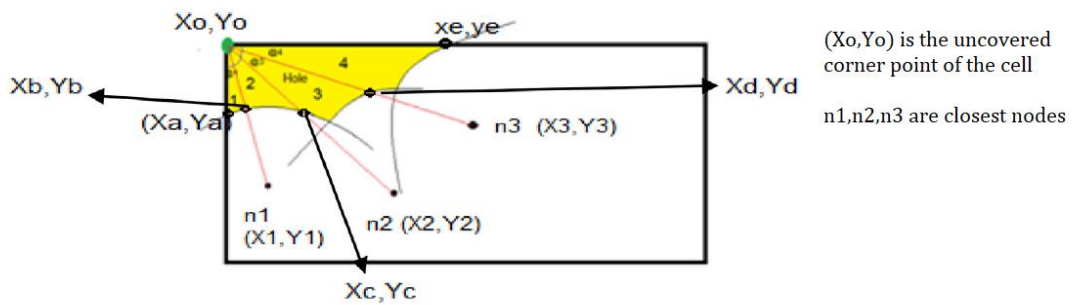

Figure 6: Hole detection using Grid algorithm. Copied from [23].

## 6.0 Hole recovery:

In [26, 27, 20, 28, 29] , some computational geometry approaches were used to heal coverage holes in WSN. In this paper we will highlight the recent works regarding computational geometry approach for healing the holes in WSN.

Kang [20] proposed algorithm that starts by finding the boundary critical points and Removing Redundant ones. Then, it selects the nearest two neighbour points to guarantee the new node will cover maximum part of uncovered area. After that it calculates the location on the new node and deploys it. It repeats this steps until the boundary critical point list is finished. Each node is able to find its boundary critical points from its neighbour information only. The time complexity of the algorithm depends on number of deployed sensors, number of holes, and the holes size. Finding boundary critical point need $O(d*n)$, $d<= n$, where d is the maximum number of neighbouring sensors.  Selecting the nearest neighbour and calculating the location need $O(2m)$, $m<=n$, where the m is the maximum number of boundary critical points. The overall complexity is $O(mdn)$. This algorithm introduces a good coverage strategy by using the boundary critical points, however, it has a large time complexity, It is more efficient with known nodes coordinates. Furthermore, adding a node to cover every adjacent boundary nodes produce high degree of overlapping. Furthermore, it neglected non-convex hull case.

After that, Aliouane and Benchaiba [28] algorithm was proposed with the similar idea but it takes overlapping in the consideration. The algorithm first find the centre point cp of a circle intersect two succeeding boundary points bp, so the region around them will be covered. The algorithm involves all boundary points to specify the appropriate location of the centre points to cover the holes.  The algorithm begins by initiating boundary node to find the centre point of a circle (with radius Rs) that has the two boundary points, see Figure 7.

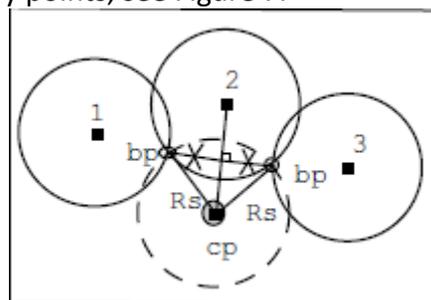

Figure 7: Location of a new node [28]

The rest of boundary nodes find the rest centre points' locations. At this point all the holes will be covered but this require a great number of nodes and also result high overlapping area, see figure 8. To recover those weakness points the algorithm uses two points to find the centre point Cp, the new intersection point np between the origin circle and the new circle and uncovered boundary point. The new intersection point is used to find the next centre point. After the next boundary node deliver the location of centre point from its neighbour, it performs the same steps using the new intersection point np to locate the centre point cp.

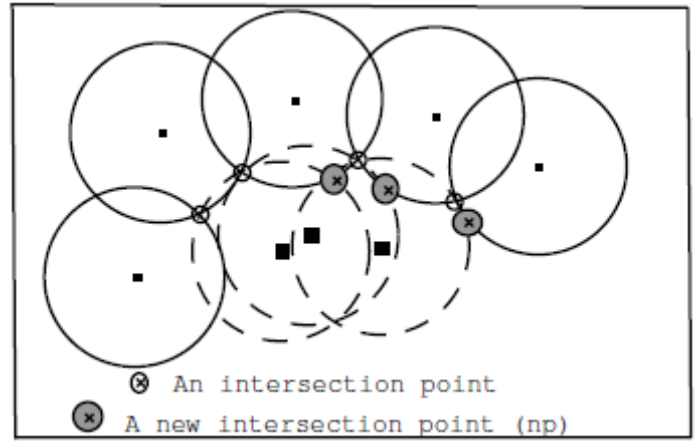
Figure 8: Large Overlapping [28]

Three possible cases may be appear: first, the new intersection point np cross one of the boundary points, so the centre can be determine by using the two boundary points. Second case, new intersection point is under the sensing area of the neighbour circle, so the first boundary point is inside the new circle. Then, The sensing neighbour use the new intersection point and its uncovered boundary point to discover the centre of circle cp. Third case, the two boundary points of the sensing neighbour are under the new circle, here no need to find the centre point. The algorithm is applied with different number of randomly deployed nodes with. The sensing 10m and 20m sensing range and communication range respectively. The results proof that the algorithm 100% heal the coverage holes. However, it is clear that the algorithm requires a lot of procedure which make the time complexity too much.

*Verma and Sharma* [29] *proposed algorithm starts by selecting one* Intersection Points then compute the Euclidian distant from this point to the others intersection points in clockwise. Then, it find the maximum Euclidian distance and find two points N1 and N2 where the N1Ci = N1Cj = Rs & N2Ci = N2Cj= Rs , CI and CJ are the two intersection points that have the maximum Euclidian distant, Rs is the sensing radius. There are three cases: Firstly, N1 and N2 already covered then deploy new node Sn on Midpoint of bibj. Secondly, N1 is covered and N2 is not covered then deploy new node on N2. The third case N1 is uncovered then the new sensor on N1. This algorithm requires less number of healing sensors. It also works for both convex and non-convex polygons. However, it is complex algorithm that require calculating the un covered area, intersection points, specifying if the area is open or closed, and calculating the Euclidian distance from one intersection point to all other intersection points.

It is clear that not all the articles mentioned to the time complexity which make it difficulties to compare between those algorithms accurately. The best thing to do is to build those algorithms using a simulation program to evaluate and compare their performances. Verma and Sharma [29] algorithms produce less overlapping and use less number of nodes to heal the holes, but there is still overlapping which is not efficient according to quality of service.

**7.0 Conclusion and future works**

This paper has discussed the hole detecting and recovery in homogeneous area WSN based on computational geometry. It binds the most recent solutions for the problem in one article by reviewing number of proposed algorithms which try to solve the problem. This article provides the researchers by the recent proposed solutions for the mentioned problem. It also tries to bring their attentions to computational geometry as one of the computer science fields that can be included to solve WSN challenges. It is clear that there are a strong relation between sensing range, communication range, hole detection, and hole healing. Nevertheless a few number of articles mentioned the time complexity for the proposed algorithms, which make it difficult to compare between them.

For the future work, we will try to implement the mentioned algorithms using a simulation programs such as MatLab. We will also include the non-computational geometry algorithms that tried to address the problem to add a section that compare between computational geometry and un-computational geometry algorithms. After that we will be able to submit it to a journal.